\documentclass[journal,twoside,web]{ieeecolor}
\usepackage{generic}
\usepackage{cite}
\usepackage{amsmath,amssymb,amsfonts}
\usepackage{algorithmic}
\usepackage{graphicx}
\usepackage{algorithm,algorithmic}
\usepackage{hyperref}
\usepackage{textcomp}
\usepackage[caption=false,font=footnotesize]{subfig}  

\def\BibTeX{{\rm B\kern-.05em{\sc i\kern-.025em b}\kern-.08em
    T\kern-.1667em\lower.7ex\hbox{E}\kern-.125emX}}
\begin{document}
\title{Estimating Visceral Adiposity from Wrist-Worn Accelerometry}
\author{James R. Williamson, Andrew Alini, Brian A. Telfer, \IEEEmembership{Senior Member, IEEE}, Adam W. Potter, Karl E. Friedl
\thanks{DISTRIBUTION STATEMENT A. Approved for public release. Distribution is unlimited.
This material is based upon work supported by the Department of the Army under Air Force Contract No. FA8702-15-D-0001. Any opinions, findings, conclusions or recommendations expressed in this material are those of the author(s) and do not necessarily reflect the views of the Department of the Army.\newline
\newline
\copyright ~2025 Massachusetts Institute of Technology.
\newline
\newline 
Delivered to the U.S. Government with Unlimited Rights, as defined in DFARS Part 252.227-7013 or 7014 (Feb 2014). Notwithstanding any copyright notice, U.S. Government rights in this work are defined by DFARS 252.227-7013 or DFARS 252.227-7014 as detailed above. Use of this work other than as specifically authorized by the U.S. Government may violate any copyrights that exist in this work.
}
\thanks{This project was supported by funds provided for the “modernization of Defense Department readiness standards research initiative” contracted through the USAF. 
The accelerometry data component of the 
NHANES project was funded through grants from the US Army Telemedicine and Advanced Technology Research Center, Fort Detrick, Maryland, 2011-2013.} 
\thanks{This work has been submitted to the IEEE for possible publication. Copyright may be transferred without notice, after which this version may no longer be accessible.}
\thanks{James R. Williamson is with the Lincoln Laboratory, Massachusetts Institute of Technology, Lexington, MA 02421, USA.}
\thanks{Andrew Alini is with the Lincoln Laboratory, Massachusetts Institute of Technology, Lexington, MA 02421, USA.}
\thanks{Brian A. Telfer is with the Lincoln Laboratory, Massachusetts Institute of Technology, Lexington, MA 02421, USA.}
\thanks{Adam W. Potter is with the U.S. Army Research Institute of Environmental Medicine, Natick, MA 01760, USA}
\thanks{Karl E. Friedl is with the U.S. Army Research Institute of Environmental Medicine, Natick, MA 01760, USA}
}

\maketitle

\begin{abstract}
Visceral adipose tissue (VAT) is a key marker of both metabolic health and habitual physical activity (PA).  Excess VAT is highly correlated with type 2 diabetes and insulin resistance. The mechanistic basis for this pathophysiology relates to overloading the liver with fatty acids. VAT is also a highly labile fat depot, with increased turnover stimulated by catecholamines during exercise.  VAT can be measured with sophisticated imaging technologies, but can also be inferred directly from PA.  We tested this relationship using National Health and Nutrition Examination Survey (NHANES) data from 2011-2014, for individuals aged 20-60 years with 7 days of accelerometry data (n=2,456 men; 2,427 women) \cite{nhanes}.  Two approaches were used for estimating VAT from activity. The first used engineered features based on movements during gait and sleep, and then ridge regression to map summary statistics of these features into a VAT estimate. The second approach used deep neural networks trained on 24 hours of continuous accelerometry. A foundation model first mapped each 10 s frame into a high-dimensional feature vector. A transformer model then mapped each day's feature vector time series into a VAT estimate, which were averaged over multiple days. For both approaches, the most accurate estimates were obtained with the addition of covariate information about subject demographics and body measurements. The best performance was obtained by combining the two approaches, resulting in VAT estimates with correlations of r=0.86. These findings demonstrate a strong relationship between PA and VAT and, by extension, between PA and metabolic health risks.
\end{abstract}


\section{Introduction}
\label{sec:introduction}

Visceral adipose tissue (VAT) is a key determinant of human metabolic health. VAT accumulation is a hallmark of aging in humans, and is a major risk factor for insulin resistance and type 2 diabetes 
\cite{wajchenberg2000subcutaneous}. 
This is explained by the effect of high concentrations of fatty acids from abdominal fat flooding the liver to change the dynamics of blood glucose and insulin regulation \cite{astrup2011role,bjorntorp1996regulation}. 
High VAT not only raises the risk for metabolic disorders, but also for
cognitive decline, Alzheimer’s disease, and disability \cite{nguyen2022potential,jain2023visceral}. 

Despite its critical role in disease pathogenesis, assessing VAT in clinical and research settings remains challenging due to the
invasive and costly nature of direct measurement techniques~\cite{maskarinec2022subcutaneous}.
VAT cannot be accurately estimated from simple body measurements such as waist circumference, waist-to-height ratio, percent body fat, and body mass index
(BMI) \cite{swainson2017prediction}.
While CT is the ideal reference, more practical dual energy X-ray absorptiometry (DXA) technology now provides a good estimate of VAT from either of the two commonly used scanner manufacturers \cite{kaul2012dual,rothney2013abdominal,bennett2023standardization}.  
VAT is more important than total body fat in metabolic health risk prediction and its relationship to physical activity may help explain the basis of the ”fit fat” concept where individuals with good fitness habits can carry higher total fat and still demonstrate reduced negative health outcomes such as coronary heart disease \cite{ortega2016obesity}. At least one previous study, using a large dataset from the Framingham Heart Study, reported that high daily moderate-to-vigorous physical activity (MVPA) measured by a hip-worn step counter was associated with a lower VAT volume~\cite{murabito2015moderate}.  This raises the possibility of using physical activity in the prediction of VAT.

Activity measurements from body-worn sensors are easy to obtain and have the potential to complement VAT estimates from body measurements. In addition, activity measures can provide useful prescriptive information about how changes in activity could lead to reductions in VAT and concomitant improvements in health.
Several large cohort studies including NHANES and UK Biobank \cite{ukbiobank} have DXA scan data and high frequency wrist-worn accelerometry in subsamples of their participants.  This study used the NHANES accelerometry cohort because of the inclusion of adults under age 40.

In this study, VAT estimation models were trained on activity variables alone, covariate variables alone, and both sets of variables jointly.
Two complementary activity analysis approaches are used. The first approach focuses on specific behaviors known to be linked to VAT, gait and sleep. Features were computed from the raw 80 Hz accelerometry data to characterize activity patterns and movement dynamics during these behaviors. Summary statistics per subject were computed from these features across all days of collected data.
The correlations of these features with VAT were analyzed to gain insight into how movements during gait and sleep were associated with VAT. 

A ridge regression model was then used to estimate VAT from the features to quantify the variance in VAT levels explained by the set of gait and sleep features.
The ability to predict VAT using the gait and sleep features was also compared to a previous approach, in which features extracted from one-minute acceleration statistics rather than raw accelerometry were used to estimate BMI~\cite{williamson2023daily}. This was done to illustrate the benefit of using raw accelerometry for extracting activity features.

The second more data agnostic approach used advanced deep learning techniques to holistically make use of the time course of accelerometry data throughout an entire day. The deep learning models mapped daily measurements into a VAT estimate, and then aggregated these estimates over multiple days. The mapping of a single day's accelerations into VAT was done using a two-step process. First, a self-supervised leaning model was used to extract general high dimensional features from each 10 s of data. Second, a transformer model was used to map the time series of these feature vectors from each day into a VAT estimate. Finally, the per day estimates were averaged across days.

VAT estimates from the two modeling approaches were obtained with and without the addition of covariate variables of demography and body measurements during training.
Fused estimates that combine the two approaches were made. Accuracy of the VAT estimates were quantified in normal, overweight, and obese subpopulations in order to determine the added benefit of using activity features jointly with covariates.

This paper provides novel contributions to the study of how activity contributes to health outcomes by predicting VAT from multiple detailed activity assessments. This was done using two alternative approaches. The first approach extracted movement features during gait and sleep and demonstrated the effectiveness of using time-delay embedding for modeling movement dynamics. The second deep learning approach combined a foundation model characterizing acceleration locally in time with a transformer network that estimated VAT based on the time series of foundation model features. The paper also demonstrated the value of using accelerometry data at high time resolution over coarser resolutions as has typically been done by researchers.
It further highlighted the improvement in accuracy of activity-based VAT estimates using demographic and body measurement information.


\section{Related Work}

\subsection{Relationship of gait to VAT}
\label{sec:gait_related_work}

Activity energy expenditure, largely composed of voluntary ambulation, has strong correlations with obesity in general, and VAT in particular. Several gait parameters correlate with obesity and VAT and motivated many of the features used in the VAT estimation algorithm developed in this paper.

Intensity of gait activity is associated with obesity and VAT. One way to measure gait intensity is via cadence or step durations. These measures are significant indicators of obesity \cite{tudor2018fast}, \cite{scataglini2024difference}, and cadence levels of at least 100 steps per minute have been recommended for reducing obesity \cite{tudor2018fast}.

Another way to measure gait intensity is with the magnitude of accelerations during gait.
Firstly, wrist-worn accelerations can be mapped with reasonable accuracy to gait speed, from slow walking up to running \cite{dawkins2022normative}. Secondly, high intensity gait (and aerobic exercise in general) is beneficial for reducing VAT \cite{chang2021effect}, \cite{vissers2013effect}, \cite{murabito2015moderate}.

In addition to gait intensity, the distribution of gait over time is also relevant. This includes the total amount of gait, total gait intensity, and the distribution of gait intensities over time. 
Duration of walking time is negatively associated with BMI and VAT, whereas sedentary time is positively associated.
\cite{ando2020association}, 
\cite{riechman2002association}
\cite{belavy2014preferential}.
Additionally, in adults above the age of 65, the total amount of daily activity was not associated with increased mortality risk, whereas activity fragmentation was positively associated with mortality risk \cite{wanigatunga2019association}. 
Activity fragmentation refers to a reduced consistency in maintaining moderate activity, resulting in more frequent shifts between moderate and low activity levels.

\subsection{Relationship of sleep to VAT}

Sleep duration and quality have also been associated with VAT.
First, short sleep duration is a risk factor for VAT accumulation \cite{yu2022sleep}, and improvements from short to adequate sleep duration over a six year period have been associated with a long-term reduction in VAT \cite{chaput2014change}.
Second, visceral adiposity has been linked to obstructive sleep apnea and poor health outcomes such as metabolic syndrome \cite{bozkurt2016visceral}
\cite{zou2020use},
\cite{vgontzas2008does},
\cite{chen2016applicability}.
Accelerometry has been used to evaluate the causal relationship between sleep fragmentation parameters measured in free-living conditions and general and abdominal obesity \cite{xu2024association}. The extracted parameters included indices of sleep fragmentation, movement, waking after sleep onset, and sleep efficiency.

\subsection{Deep learning foundation models} As deep learning has matured, foundational models for particular modalities have become more prevalent. Foundation models broadly characterize a particular data modality. They are typically trained with a large dataset using self-supervised techniques. In self-supervised learning, models learn features from an unlabeled dataset by solving a predefined task. Contrastive learning is a particularly popular technique for training foundation models, where models learn features by distinguishing similar pairs (positive) from dissimilar pairs (negative) \cite{contrast,clip}. Positive pairs are typically augmentations of the same datapoint while negative pairs are different datapoints. Once trained, foundation models can be used to extract features from high-dimensional data for downstream tasks \cite{bommasani2021opportunities}.

Yuan et al. utilized contrastive learning to train a wrist-worn accelerometry foundation model with data sampled at 30 Hz in a 10-second window \cite{oxwearable}. This model was trained using a one-dimensional residual network (1D-ResNet) based on convolutional neural networks (CNNs) \cite{cnn_paper,resnet_paper}. CNNs are a deep learning architecture that leverages convolutions in signal processing to extract hierarchical features from data. ResNets uses skip connections to enable information to bypass intermediate neural network layers, helping mitigate issues such as vanishing gradients and preserving important features. This foundation model can be used to extract features for downstream tasks.

\subsection{Deep learning transformers} Transformers are a type of neural network for sequential data that uses an attention mechanism to learn each time-step's context by analyzing its relationship across the entire sequence, known as the context window \cite{attention_paper}. With enough data, they are highly successful for learning features of sequential data and have achieved inspiring results across a multitude of tasks \cite{vit_paper,gong2021ast,zhao2023survey}. However, traditional transformers require quadratic complexity to compare each time-step to the entire context window, leading to computational complexity challenges for very long sequences. To solve this issue, various papers in natural language processing have introduced more advanced attention mechanisms to handle longer context windows \cite{longformer,reformer,linformer}. Performers are a particularly efficient transformer with an advanced attention mechanism that uses the kernel trick to approximate comparisons between each step and the entire context window, thus enabling linear computational complexity for analysis of longer sequences \cite{performer}. After features are extracted from a transformer, they are often passed into a small multi-layer perception (MLP), a simple network of nonlinear transformations and activation functions, to map features to predict the specific task \cite{wen2022transformers,khan2022transformers}.

\section{Methods}

\subsection{Data set}

NHANES periodically samples the US population for health risk factors and outcomes. NHANES participants provided written informed consent before testing and agreed to the public deidentified data sharing.  The NHANES study was approved by the National Center for Health Statistics Research Ethics Review Board \cite{nhanes,ahluwalia2016update}.  Processed activity data were obtained from the 2011-2014
NHANES data set. A subset of participants wore an accelerometer
(ActiGraph, Model GT3X+, Pensacola, FL) on their nondominant
wrist 24 h/day for at least seven consecutive
days. 
The accelerometer collected data at 80 Hz, and the axes were oriented as follows. The $x$ axis was along the length of the arm, the $y$ axis was in the forward-backward direction (in the direction of arm swing during walking), and the $z$ axis was lateral (side-to-side) \cite{straczkiewicz2019placement}. 
Previous work has demonstrated a high degree of similarity in accelerometry data among Actigraph, Axivity, and GENEActiv wrist-worn devices across both static and dynamic conditions, supporting the potential interchangeability of these sensors for gait and movement analyses \cite{williamson2019comparison}.

Subjects with ages between 20 and 60 years were
included in this analysis, with roughly equal numbers in their 20s, 30s, 40s and 50s (1,157, 1,205, 1,286, 1,235). %
This age range represents a life cycle period of stable metabolic rate \cite{pontzer2021daily}.  Subject inclusion also required the availability of
measurements of height, body weight, waist circumference, and
VAT. Waist circumference was defined as the horizontal measure around a standing person at the level of the medial iliac crest \cite{nhanes}.
The VAT measurements were
obtained using dual-energy X-ray absorptiometry (QDR 4500A fan beam densitometer, Hologics, Waltham, MA), and VAT mass was recorded in grams. VAT was defined as a rectangular area within the torso, 5 cm above the line drawn horizontally at the level of the iliac crest. 
Subcutaneous fat was estimated by assessment of the subcutaneous fat layer at the lateral margins of the torso and subtracted from the estimate \cite{bennett2023standardization}. Other selection criteria included at least three hours total duration of detected gait.
The data set comprised 4,883 subjects (2,456
male; 2,427 female) who met all the selection criteria.
Table~\ref{table:subject_measurements} shows the means and standard deviations of demographic and body measurements in the data set.

\begin{table}
\caption{Subject demographics and measurements}
\begin{tabular}{|l|l|l|l|}
\hline
Measurement & All & Male & Female \\
\hline
Age (y)     &   39.7 $\pm$ 11.4    &  39.2 $\pm$ 11.4        & 40.1 $\pm$ 11.4          \\
Weight (kg)  &  82.3 $\pm$ 21.5    & 87.3  $\pm$ 20.8    & 77.3  $\pm$ 20.9       \\
Height (c)  &  168.5  $\pm$ 9.8   & 175.1  $\pm$ 7.4    & 161.8    $\pm$ 7.1       \\
Waist (c)   &  96.9 $\pm$ 18.2    & 98.4  $\pm$ 17.1   & 95.4  $\pm$    19.1    \\
BMI (kg/m2) & 28.9  $\pm$ 6.9    & 28.4  $\pm$ 6.2   & 29.5  $\pm$ 7.5       \\
VAT (g)     & 496.0  $\pm$ 273.5   & 514.0  $\pm$ 266.7   &  477.8 $\pm$ 279.1        \\
\hline
\end{tabular}
\label{table:subject_measurements}
\end{table}

\subsection{Gait features}

Gait feature analysis relies on detection of local gait data frames, which contain a sufficient level of acceleration magnitude and appropriate periodicity to be consistent with gait behavior. Motivated by previous research described in Section~\ref{sec:gait_related_work}, the frame-based gait features were organized into different classes: 1) cadence and periodicity features, 2) movement intensity features, and 3) temporal pattern features. An additional fourth class of gait dynamics features was also included. This analyzed time-frequency and distributional properties of acceleration measurements, independent of absolute acceleration magnitudes.

\subsubsection{Detecting gait frames}
\label{sec:detect_gait_frames}

Following an approach used in previous work \cite{williamson2021parkinsons},\cite{buller2022gait},\cite{williamson2022using}, gait bouts were segmented from accelerometry data based on average local activity level. 
This is defined as the acceleration magnitude standard deviation (MSD).
At time $t$, the acceleration magnitude is
\begin{eqnarray}
m(t) = \sqrt{x(t)^2 + y(t)^2 + z(t)^2}.
\end{eqnarray}
The standard deviation of $m(t)$ is denoted by $\sigma_{m(t)}$, which is computed over a 10~s window, centered at $t$.
Gait bouts were segmented based on time periods in which $\sigma_{m(t)} > 0.1$ (g), across a continuous interval of at least 10~s duration. A single bout could also include one or more subthreshold gaps, provided the gaps had durations of 15~s or less.

\begin{figure}[htbp]
    \centering

    \includegraphics[width=\columnwidth]{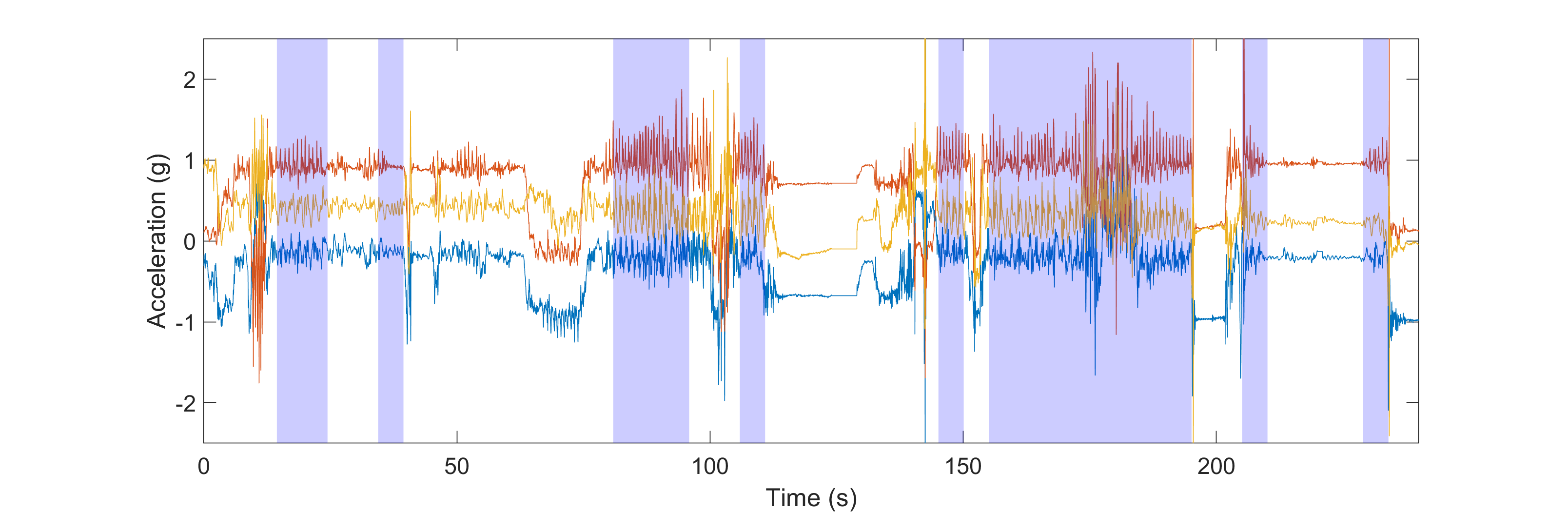}

     \subfloat[]{\includegraphics[width=0.48\columnwidth]{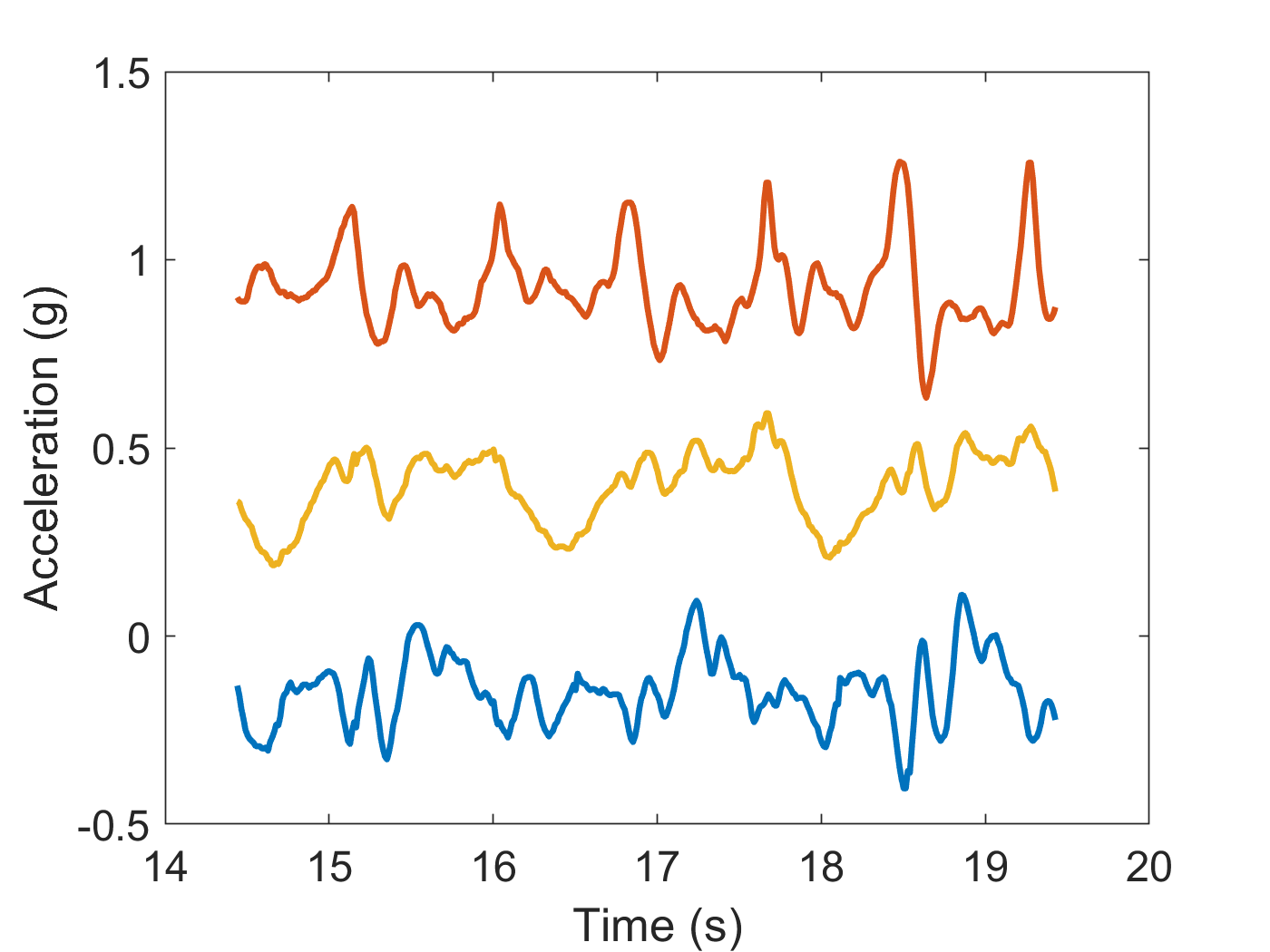}}
     \subfloat[]{\includegraphics[width=0.48\columnwidth]{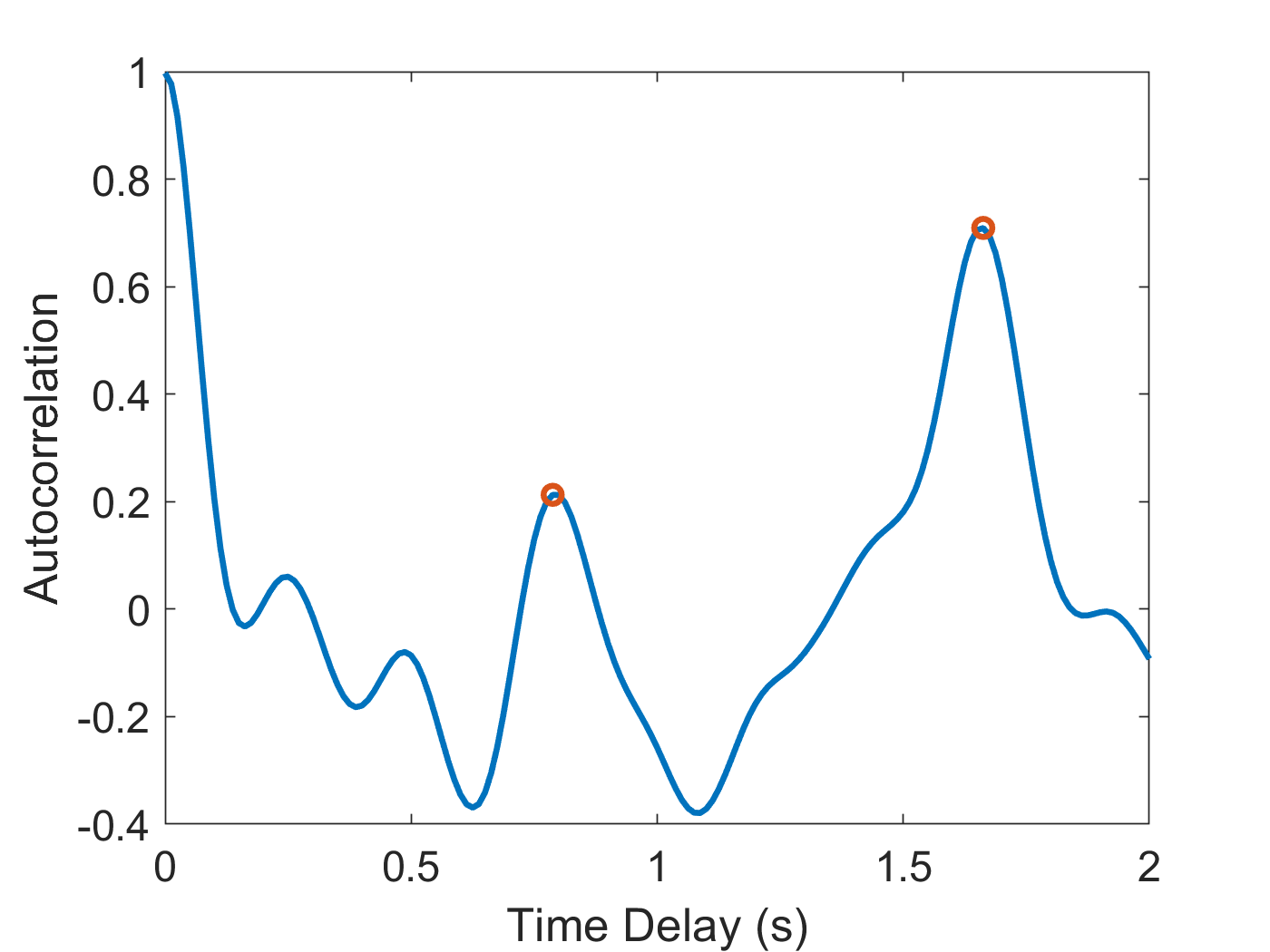}}

    \caption{
    Top: 240 s segment of acceleration data with shading indicating detected gait frames. Bottom left: acceleration from first gait frame. Bottom right: autocorrelation function for first gait frame.
    }
    \label{fig:timeline_examp}
\end{figure}

Gait bouts were divided into contiguous 5~s frames. Those frames that met a gait periodicity test were labeled as gait frames and were subsequently the source of all gait features. The gait periodicity test used the autocorrelation function of the first principal component of the acceleration data within the frame.
The periodicity test required at least one  autocorrelation peak in a plausible range of step and stride time delays between 0.35~s and 1.70~s with a peak prominence $>$ 0.2 and peak height $>$ 0.01.
Figure~\ref{fig:timeline_examp} (top) illustrates a 240 s segment of acceleration data from one subject, in which 18 gait frames were detected (shaded regions). The acceleration data from the first of these gait frames, occurring from 14.4 s to 19.4 s of this segment, is shown on the bottom left. The autocorrelation function of the first principal component of the acceleration data is shown on the bottom right. There are two peaks that meet the gait detection criteria, representing an average step duration of 0.79 s and an average stride duration of 1.66 s.  
Similar methods of gait detection have been used previously to measure movement abnormalities due to Parkinson's disease from wrist-worn accelerometry~\cite{williamson2021parkinsons}, to predict exertional heat stroke from torso accelerometry~\cite{buller2022gait}, and to detect the behavioral effects of blast exposure from head-worn accelerometry~\cite{williamson2022using}. 

\subsubsection{Cadence and periodicity}
Gait cadence and periodicity features
were computed based on the autocorrelation peak used to detect the existence of gait. The peak height indicates the level of periodicity during the 5~s frame. The time delay of the autocorrelation peak indicates the average step (or possibly stride) duration during the frame. The minimum and maximum step durations were defined as 0.35~s and 0.85~s. Due to the possibility that the peak could represent a stride (double step) duration, the maximum threshold was doubled to 1.7~s. This was done following an internal validation of the algorithm on a data set which contained concurrent wrist and hip accelerations \cite{karas2021labeled}. It was observed that wrist-worn accelerations provided accurate gait detections compared to those obtained from hip accelerations, but with occasional frequency halving due to detections of stride rather than step periodicity, and therefore required a doubling of the maximum duration threshold.
As a result, in addition to the features of autocorrelation peak height ({\em step periodicity}) and delay ({\em step duration1}), an additional step duration feature was included, in which the autocorrelation delay was halved if it was $>$ 0.85~s ({\em step duration2}).
 
\subsubsection{Movement intensity}
Acceleration MSD, which was used as a criterion for detecting gait frames in Section~\ref{sec:detect_gait_frames}, was also computed within 5~s frames as a gait movement intensity feature. Another movement intensity feature is path length (PL), which is the summed distance between successive $x,y,z$ values in each frame. The third movement intensity feature is mean absolute distance (MAD) in $x,y,z$ space, which represents the average distance between all pairs of acceleration values within a frame \cite{williamson2021parkinsons},\cite{buller2022gait}.

\subsubsection{Gait patterns over time}
Gait duration represented the amount of gait behavior per subject and was computed by counting the total number of gait frames across all days. A related feature is cumulative gait intensity, which was computed by summing the MSD values across all gait frames.
MSD was also the basis for features that represent the joint probability distribution of gait intensity levels across successive frames. These features were used due to previous findings that poor health is associated with activity fragmentation, marked by a lower incidence of sustained high or moderate activity levels, and a greater incidence of transitions into low activity gait levels~\cite{wanigatunga2019association}.

The gait intensity transition probabilities were computed as follows. Each gait frame was assigned one of 4 activity levels based on its MSD value. These bins represent gait of very-low, low, moderate, and high intensity. The bin delimiters were selected by associating MSD values with a similar activity level measure, which is the Euclidean norm minus one (ENMO). ENMO was used in \cite{dawkins2022normative} to empirically map activity levels from wrist-worn accelerometry into different walking speeds. Based on this association, the bins for categorizing MSD values correspond to the following walking speeds: 1) very slow walking ($\leq$ 0.125~g); 2) slow walking ($>$ 0.125~g and $\leq$ 0.375~g); 3) moderate and fast walking ($>$ 0.375 g and $\leq$ 1.0~g); 4) running ($>$ 1.0~g).
A $4 \times 4$ joint probability distribution was created, containing the joint probabilities of all transitions over successive frames between the four gait intensity categories.
Note that MSD was adopted as a basic gait intensity measure over ENMO because in an internal analysis it was found to have a higher correlation with VAT and greater predictive value as a feature in regression modeling.

\subsubsection{Movement dynamics}
Additional gait features were computed that represent distributional and time-frequency properties of gait, but with acceleration amplitude factored out. In other words, the dynamics features make use only of z-scored acceleration time series within each frame. These features include PL and MAD. MAD is computed in 3-d space with values outside 4 standard deviations removed. MAD is also computed within six 2-d axis pairs: $xx$, $xy$, $xz$, $yy$, $yz$, $zz$, with values outside 2 standard deviations removed, as in \cite{williamson2021parkinsons},\cite{buller2022gait}.

The most important movement dynamics features were based on a general approach that quantifies movement dimensionality and time-frequency properties based on the structure of correlations among the three acceleration axes at multiple time delays and at multiple delay scales \cite{williamson2021parkinsons},\cite{williamson2022using}. 
These correlation structure features are represented by the eigenspectra of high dimensional matrices constructed with  time-delay embedding (TDE). This technique was described in detail in  \cite{williamson2021parkinsons} and \cite{williamson2022using}. Four delay scales were used, with 7 delays per scale. 
There were thus 21 eigenvalue features per scale (7 delays times 3 acceleration axes), and a total of 84 features across the 4 scales.
The scale-depending delay spacings were $\{3,7,15,31\}$, which correspond to delays of $\{37.5,87.5,187.5,387.5\}$ ms given that the acceleration data was sampled at 80 Hz. 

\subsubsection{Summary statistics}

The frame-level measures of gait described above were aggregated per subject to produce summary statistics of mean and standard deviation. These were used as inputs to a ridge regression model for estimating VAT. Table~\ref{table:gait_features} lists the gait features and the dimensionality of their summary statistics. For each frame level feature there are two summary statistics features (mean and std). The global correlations of the summary statistics  with VAT across all 4,883 subjects are also listed, which will be discussed in Section~\ref{sec:Results}.

\begin{table}
\caption{Gait features and VAT correlations}
\begin{tabular}{|l|l|l|l|}
\hline
Property & Features & Dim. & VAT r \\
\hline
Cadence and     & Step duration1             & 1$\times$2    & ~0.24 -0.02\\
periodicity     & Step duration2       & 1$\times$2    & ~0.22 ~0.07 \\
                & Step periodicity   & 1$\times$2    &-0.15 -0.17 \\
\hline
Movement & Accel MSD     & 1$\times$2 & -0.28 -0.26\\
intensity     & Accel PL      & 1$\times$2 & -0.25 -0.18\\
(raw accel)                & Accel MAD     & 1$\times$2 & -0.22 -0.24\\
\hline
Gait patterns   & Intensity     & 4$\times$4 & ~0.21 ~0.25 -0.05 -0.20 \\
over time       & transition    & & ~0.24 ~0.04 -0.16 -0.25 \\
                & probabilities & & -0.04 -0.16 -0.21 -0.26\\
                & & & -0.18 -0.25 -0.27 -0.29 \\
                & Frame count   & 1$\times$1 & -0.09 \\
                & Frame MSD sum & 1$\times$1 & -0.16 \\

\hline
Movement   & Accel PL      & 1$\times$2& -0.18 ~0.18 \\
dynamics    & Accel MAD     & 1$\times$2& -0.06 ~0.12\\
(z-scored                & Accel MAD xx  & 1$\times$2& -0.08 ~0.10 \\
accel)                & Accel MAD xy  & 1$\times$2& -0.10 ~0.06 \\
                & Accel MAD xz  & 1$\times$2& -0.02 ~0.15 \\
                & Accel MAD yy  & 1$\times$2& -0.17 ~0.08 \\
                & Accel MAD yz  & 1$\times$2& -0.05 ~0.15 \\
                & Accel MAD zz  & 1$\times$2& ~0.02 ~0.21 \\
                & TDE eigenvalue & 4$\times$21& eigval rank 4: \\ 
                & means, 4 scales & & ~0.14 ~0.20 ~0.40 ~0.34 \\
                & TDE eigenvalue  & 4$\times$21& eigval rank 4: \\
                & stds, 4 scales & & ~0.26 ~0.07 ~0.02 -0.06\\
\hline
\end{tabular}
\label{table:gait_features}
\end{table}

\subsection{Sleep features}

Sleep feature analysis relies on the automated labeling of sleep bouts, which are of prolonged time periods with predominantly low activity levels. Within sleep bouts, there are alternations of short bursts of detected movements, interspersed with longer intervals of inactivity. Overall sleep quality was characterized by sleep fragmentation features. These represent the distribution of timings of the movement and inter-movement intervals. Sleep feature analysis also included the analysis of the detected movement segments via the subsequent extraction of frame-level movement features, using many of the same techniques used to analyze gait movements.

\subsubsection{Segmenting sleep bouts and sleep movements}

Following an approach used in previous work for detecting prolonged periods of low movement\cite{williamson2021parkinsons},\cite{williamson2022using}, sleep bouts were segmented from accelerometry data based on average local activity level. 
Specifically, sleep bouts were segmented based on time periods in which $\sigma_{m(t)} < 0.01$ across an interval of at least 2 hours, but during which suprathreshold intervals of 2 minutes or less were allowed. Within sleep bouts, movement segments were detected based on time intervals with high activity levels, based on $\sigma_{m(t)} > 0.05$ over a duration of at least 10 s. These movement segments were then divided into contiguous 5 s movement frames for movement feature analysis. 

\subsubsection{Sleep fragmentation}

Sleep fragmentation features measure the distribution of sleep durations in between movements, as well as the distribution of the movement segments themselves. The time delimiters separating the 5 sleep interval durations are $\{4,8,16,32 \}$ minutes and the time delimiters separating the 5 movement interval durations were $\{12,25,100,200 \}$ seconds. The histograms of sleep and movement duration counts were normalized per subject to create probability distributions.

\subsubsection{Sleep movements}

Sleep movement features were computed within all 5~s sleep movement frames in which the frame-level MSD was greater than 0.005. The same frame-level movement intensity and dynamics features, and their summary statistics, that were used to analyze gait were also extracted for these movement features, as summarized in Table~\ref{table:sleep_features}. In addition, median accelerations in the $x$, $y$, and $z$ axes were extracted because they provide information about average wrist position with respect to the gravity vector during sleep movements. 

\begin{table}
\caption{Sleep features and VAT correlations}
\begin{tabular}{|l|l|l|l|}
\hline
Property & Features & Dim. & VAT r \\
\hline
Fragmen- & Sleep dur prob    & 1$\times$5    & ~0.11 -0.07 -0.10 -0.02 ~0.03\\
tation    & Move dur prob      & 1$\times$5    & -0.04 -0.12 -0.03 ~0.11 ~0.17 \\
\hline
Move           & Accel MSD     & 1$\times$2 & -0.12 -0.08\\
intensity      & Accel PL      & 1$\times$2 & -0.21 -0.18\\
(raw       & Accel MAD     & 1$\times$2 & -0.11 -0.12\\
accel)            & Median x  & 1$\times$2 & -0.01 ~0.04\\
                & Median y  & 1$\times$2 & ~0.24 ~0.17\\
                & Median z  & 1$\times$2 & -0.09 -0.19\\
\hline
Move            & Accel PL      & 1$\times$2& -0.05 ~0.02 \\
dynamics       & Accel MAD     & 1$\times$2& ~0.14 -0.08\\
(z-scored        & Accel MAD xx  & 1$\times$2& ~0.13 -0.06 \\
accel)          & Accel MAD xy  & 1$\times$2& ~0.14 -0.08 \\
                & Accel MAD xz  & 1$\times$2& ~0.14 -0.09 \\
                & Accel MAD yy  & 1$\times$2& ~0.12 -0.08 \\
                & Accel MAD yz  & 1$\times$2& ~0.15 -0.10 \\
                & Accel MAD zz  & 1$\times$2& ~0.13 -0.08 \\
                & TDE eigval & 4$\times$21& eigval rank 3: \\ 
                & means, 4 scales & & ~0.04 ~0.07 ~0.11 ~0.14 \\
                & TDE eigval  & 4$\times$21& eigval rank 3: \\
                & stds, 4 scales & & ~0.21 ~0.21 ~0.13 -0.09\\
\hline
\end{tabular}
\label{table:sleep_features}
\end{table}

\subsection{Covariate variables}

An additional goal of this study was to investigate the additive value of measured covariate variables in estimating  VAT from activity. In the NHANES data set, the following demographic and body measurement variables were available and useful for estimating VAT: age, gender, height, weight, BMI, and waist circumference. 
%


\subsection{Regression model based on gait, sleep, and MIMS-based features}

\begin{figure}[!t]
\centerline{\includegraphics[width=\columnwidth]{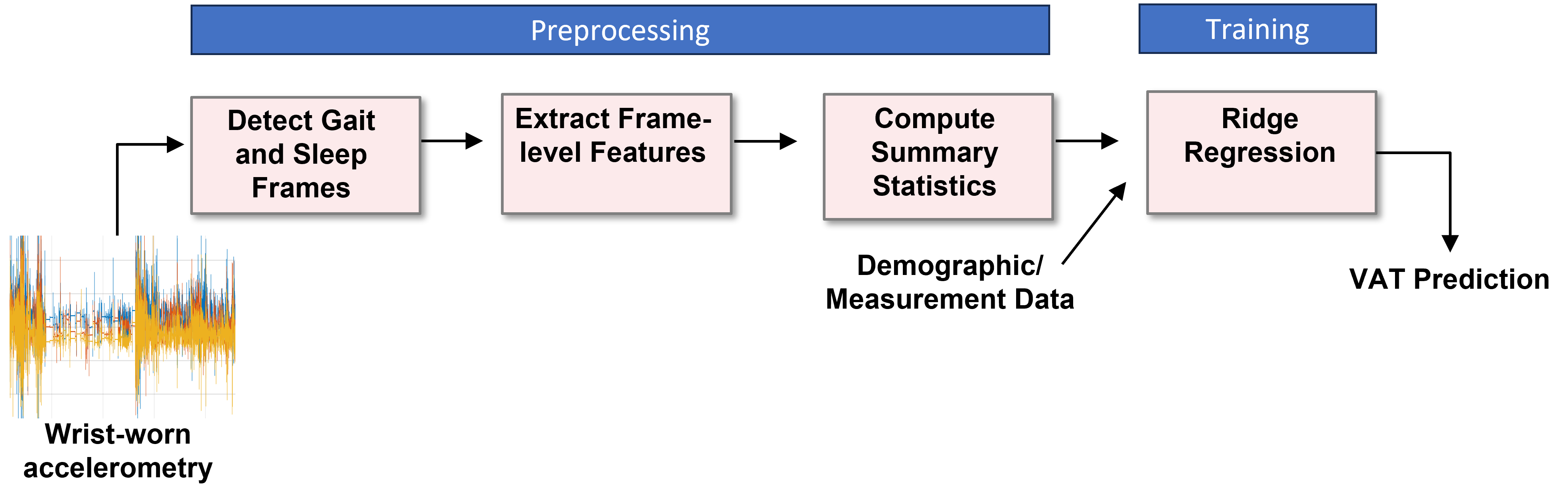}}
\caption{
Model architecture for VAT estimation based on gait and sleep features.
}
\label{fig:gait_sleep_pipeline}
\end{figure}

Gait and sleep feature correlations with VAT were analyzed to provide insight into how different types of movements are associated with VAT level. 
Ridge regression with a ridge parameter of 0.1 was then used to map concatenated feature vectors into VAT estimates, using cross validation with the data randomly divided between train and test sets with ratios of 0.8 and 0.2. This was done with 30 random data partitions and average performance on the test folds was computed.
The total dimensionality of gait features was 214 and of sleep features was 206. 
The regression model was applied with and without the addition of the 6 covariate variables that describe demographics and body measurements.
Figure~\ref{fig:gait_sleep_pipeline} diagrams the processing pipeline of gait and sleep feature extraction, followed by regression modeling.

The same procedure was also applied, for comparison, to the activity features derived from 1-minute resolution Monitor-Independent Movement Summary (MIMS) units \cite{john2019open}. In \cite{williamson2023daily}, MIMS units were the basis for features characterizing activity profiles as a function of time of day, and features characterizing activity variability over time. Due to the more restrictive subject inclusion criteria used in \cite{williamson2023daily}, this comparison was performed on a subset of the data set (3,042 subjects versus 4,883 subjects).

\subsection{Deep learning models}
\begin{figure}[!t]
\centerline{\includegraphics[width=\columnwidth]{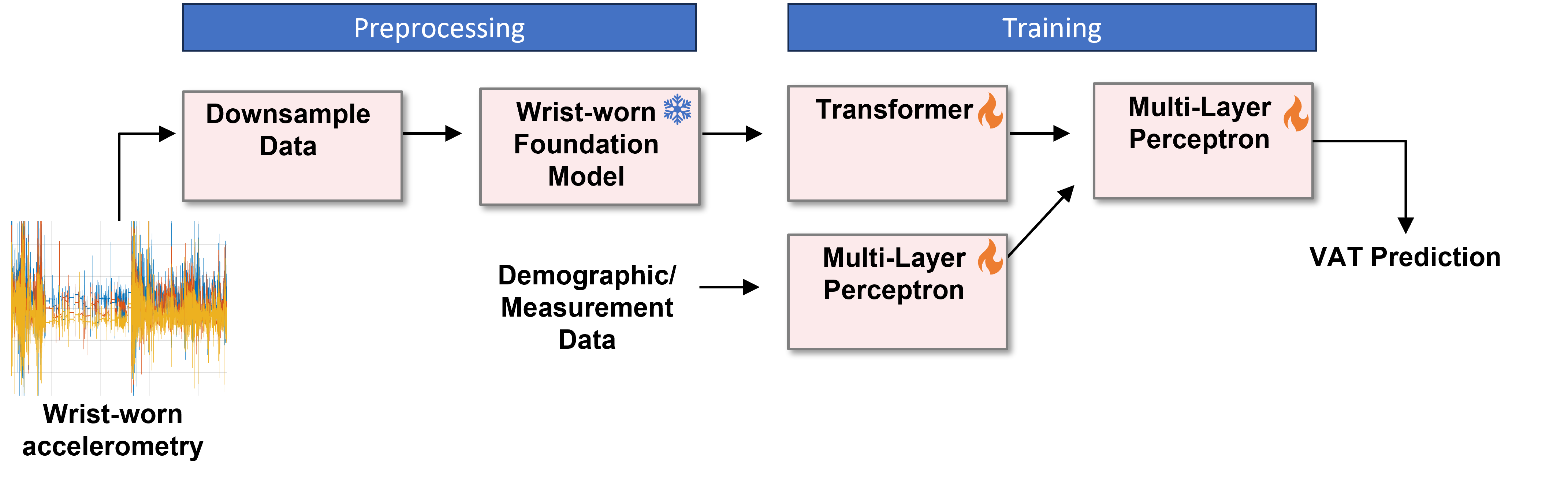}}
\caption{Deep Learning Architecture: Foundation model extracts features of accelerometry data. extracted features and other data-sources are passed into a transformer and MLP respectively. Finally a MLP is used to fuse all the data together and estimate VAT. The snowflake and fire icons indicate the model's weights were frozen for inference or were adapted for training respectively.}
\label{fig:deep_learning_arch}
\end{figure}

Deep learning models have the potential advantages of uncovering complex patterns from raw accelerometry data. Therefore, we leveraged deep neural network (DNN) architectures to predict VAT markers from wrist-worn accelerometry data and body measurement and demographic data. As shown in Figure~\ref{fig:deep_learning_arch}, the deep learning architecture can be broken down into a feature extractor for accelerometry data and a regression model for predicting VAT markers.

\subsubsection{Foundation Model for Feature Extraction of Accerometry data}  
It would be impractical to use raw accelerometry across multiple days directly as input to a transformer model. Across multiple days of data recorded at 80 Hz, the data sequence length is as high as 50 million samples for each individual.  To reduce the sequence length to a more manageable size, we used Yuan et al.'s wrist-worn accerometry foundation model to extract a feature vector for every ten seconds \cite{oxwearable}. 
To ensure distribution match, the 80 Hz accerometry data was down-sampled to 30 Hz for input to the model. Note that the foundation model was pretrained to deemphasize periods of low activity.

\subsubsection{Regression Model using MLPs and Transformers}

There are three parts of the regression model: a transformer for the extracted accelerometry features, an MLP for the covariate variables (demography and body measurements), and an MLP to fuse all the final representations together for health marker prediction. The data input into the transformer consist of a single day's worth of the 10 s extracted features from Yuan et al.'s foundation model. This provides the transformer with information about temporal activity patterns that span an entire day, which is potentially a much richer source of information than treating each  ten-second window independently, as is done when computing simple summary statistics. The data was passed into either a basic transformer with a CNN as an embedder, or a more advanced performer network. 

The body measurement and demographic data were passed into an MLP to extract representations over the metadata. Categorical data were one-hot encoded and numerical data were rescaled between zero and one before being fed into the MLP. Finally, the representations from both networks were concatenated and passed into a final MLP for VAT estimation.

\subsubsection{Training \& Implementation Details}
We trained models for both the basic and the performer transformers with the following data types: 1) accelerometry data only, 2) demographic and measurement data, and 3) accelerometry, demographic and measurement data. 

The dimensionality of the representations was 512 and 1024 for the basic and performer respectively with both having 8 heads. The basic transformer's CNN embedder had 512 output channels and a stride and kernel of 6. The stride and kernel for the basic transformer’s CNN embedder were chosen to be the same to ensure disjoint patches when inputted into the transformer. The MLP extracting information from demographic and measurement data had 3 layers with output dimensionality of 32. Finally, the fusion MLP of all the data-sources at the end of the model had four layers with sizes 512, 128, 64, and 1.  We used two V100-32GB GPUs for training and inference. 
In order to ensure the performer network could fit onto the GPUs, the basic had 6 blocks while the performer had only 1. A deeper performer network may provide improved performance.

As the NHANES dataset had undefined fields, the data were prefiltered to only include subjects with gender, height, waist circumference, and weight measurements, as well as the age range being between 20 and 60 years old. This also ensured that the various models were trained on the same data for effective comparison. We split the data into train, validation, and test with 0.7, 0.1, and 0.2 ratios respectively. We performed grid search on the batch size and learning rates using the validation set and chose the best configuration for experiments. As defined by Krizhevsky \cite{krizhevsky2014one} and Goyal et al. \cite{goyal2017accurate}, we used the learning rate scaling rule defined by the following equation:
\begin{equation}
    \centering
    \mu_{true}=\mu_{base}*\frac{B}{256},
\end{equation}
where the true learning rate ($\mu_{true}$) is a function of the base learning rate ($\mu_{base}$) and the batch size (B). More detailed training information can be found in Table \ref{table:deep_learning_config}.

\begin{table}
\centering
\caption{Training Configuration Details}
\begin{tabular}{|l|l|}
\hline
\multicolumn{2}{|l|}{Training Hyperparameters}\\
\hline
epochs & 30  \\
warm-up epochs & 1  \\
warm-up step-size & 0.1\\
base learning rate& 0.0005 \\
loss & MAE\\
optimizer & AdamW \cite{adamW}\\
momentum & 0.9\\
scheduler & cosine decay\\
batch size &32\\
dropout & 0.1\\
seed & 42\\

\hline
\end{tabular}
\label{table:deep_learning_config}
\end{table}


\section{Results}
\label{sec:Results}

\subsection{Correlations of gait features with VAT}

The rightmost column of Table~\ref{table:gait_features} shows the correlations of the gait summary statistics features with VAT. Frame-based features produce two summary statistics (mean and std dev), resulting in two numbers in this column. The directionality of correlations is consistent with published literature. 

Step durations were positively correlated with VAT. Step periodicity was negatively correlated, possibly due to slower gait generally being less periodic.
Average gait intensity features were negatively correlated with VAT, both in terms of mean and standard deviation. Therefore, VAT was associated not only with a lower average gait intensity, but also lower variability of intensity.

The gait transition probability correlations relate to how gait intensity varies across successive gait frames. The listed correlations consist of 4 rows and 4 columns. The first row shows correlations for transition probabilities from 1) very low intensity to: a) very low, b) low, c) moderate, and d) high intensity. The next three rows similarly show correlations from 2) low, 3) moderate, and 4) high intensities. VAT is positively correlated with transition probabilities involving very low and low intensities, and negatively correlated with transitions among moderate and high intensities. These results are directionally consistent with findings about activity fragmentation relating to poor health outcomes \cite{wanigatunga2019association}.

Total gait count and total summed gait intensity show negative correlations with VAT, as expected, but of lower magnitude than features involving average gait intensity or intensity transitions.
This is consistent with findings that activity intensity is more important than total activity duration \cite{chang2021effect}, \cite{vissers2013effect}, \cite{murabito2015moderate}.

The first class of gait dynamics features measured how acceleration values were distributed, either across all three axes locally in time (PL) or globally across the entire gait frame (MAD), with the magnitude of acceleration factored out via z-scoring. There were also measures of how the values are globally distributed along all 6 axis pairs. We see that for all of these variables, the means tend to be negatively correlated and the standard deviations  positively correlated. 
There was a moderate negative correlation between VAT and PL mean, but a positive correlation with PL standard deviation, suggesting that higher VAT is associated with lower average high-frequency acceleration but greater variability in high-frequency accelerations over time.
Along individual axis pairs, the strongest correlation is negative for MAD yy mean, which indicates VAT is associated with less dispersion in accelerations associated with forward-backward movements.  

The gait TDE eigenvalue features have the strongest correlations with VAT across all feature sets. Because these features are high dimensional (21 eigenvalues per 4 different delay scales), the correlations are listed only for the eigenvalue rank that yields the strongest correlation (rank 4), which represents intermediate time frequencies at each delay scale. Here we see very high correlations with VAT in the 3rd and 4th delay scales (r=0.40 and 0.34, respectively), for which the average time delays are 0.56~s and 1.16~s. This indicates that VAT is associated with greater complexity or dimensionality of movement in an intermediate frequency range (rank 4) at time delays (scales 3 and 4) that cover an entire step or stride period. This is consistent with the negative association found between VAT and gait periodicity, but the TDE embedding feature method is a richer way to characterize the relevant differences in dynamics than is the height of the autocorrelation peak. 
In addition, a moderately strong positive association with VAT is found for the eigenvalue std dev at the first delay scale. This indicates VAT is associated with higher variability of dynamics across frames at small time scales.

\subsection{Correlations of sleep features with VAT}

The rightmost column of Table~\ref{table:sleep_features} shows the correlations of the sleep summary statistics features with VAT. 
The first class of features is a per-subject probability distribution of time intervals between movement segments, and of time durations of movement segments themselves. VAT is positively associated with a greater incidence of short between-movement sleep intervals, and of longer durations of movement intervals. However, the size of these correlations is small.

The 3-axis movement intensity features (MSD, PL, MAD) are all negatively correlated with VAT, similar to what was found with gait intensity features, albeit with smaller correlations. 
Interestingly, the median acceleration in the $y$ axis (indicating frontward-backward movements) is positively associated with VAT, which possibly suggests a higher incidence of gait movements during sleep bouts among subjects with high VAT levels.

Unlike with gait features, there is a positive correlation between VAT and movement dynamics MAD feature means, indicating a tendency toward greater acceleration dispersion in all directions for subjects with higher VAT levels.
Finally, the strongest sleep TDE eigenvalue correlations were found at rank 3, which represents movements at an intermediate temporal frequency. Here, we see weak correlations with the eigenvalue means, but strong correlations with the standard deviations at the first two delay scales, which have mean delays of 0.11 s and 0.26 s, indicating a positive association between VAT and movement variability at an intermediate frequency (rank 3) at relatively short time scales (scales 1 and 2).

\subsection{Example feature plots}

\begin{figure}[!t]
\centerline{
\includegraphics[width=\columnwidth]{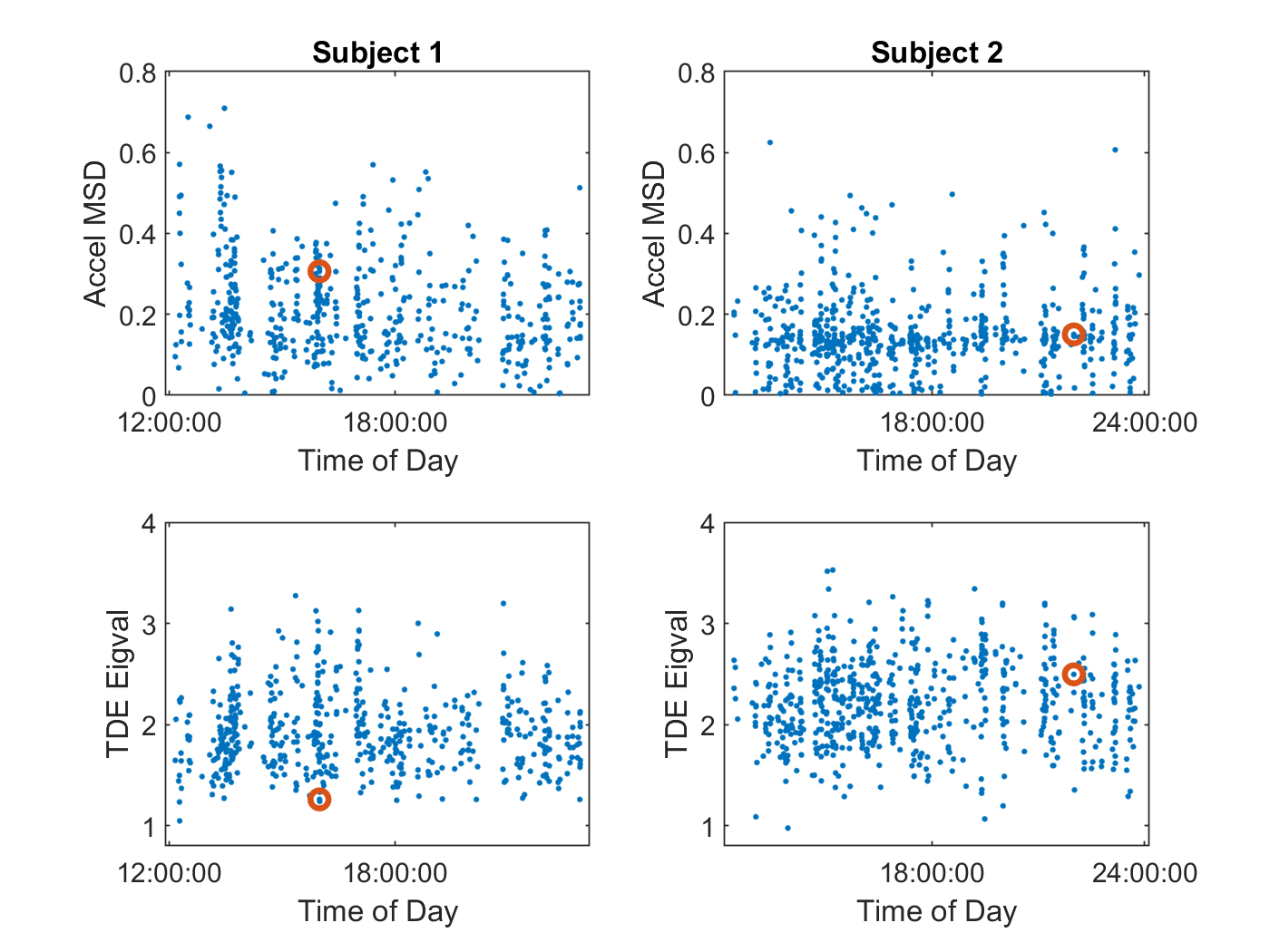}
}
\caption{Gait features from 5 s data frames based on a 12-hour period for two example subjects. Top row: gait acceleration MSD. Bottom row: gait TDE eigenvalue at scale 3 and rank 4.  
}
\label{fig:example_plot1}
\end{figure}

In Figure~\ref{fig:example_plot1}, gait features representing magnitude and dynamics were plotted for two subjects with widely differing VAT levels in order to provide a concrete illustration of how gait features discriminate VAT. Each data point was the feature value from a single 5 s gait frame plotted as a function of time of day.
Subject 1 was a 46 year old male with a BMI of 22.6 and VAT of 262.7 g. Subject 2 was a 46 year old male with a BMI of 36.2 and VAT of 1,266.4 g.
While there was an entire week of accelerometry data for each subject, the features were plotted for a 12 hour period from noon to midnight on a weekday (a time period in which both subjects maintained activity), in order to provide a closeup view. The first gait feature (top row) is gait accel MSD. The mean and standard deviation of this feature were negatively correlated with VAT over the entire data set (r=-0.28 and -0.26; see Table~\ref{table:gait_features}). The second gait feature was the TDE eigenvalue at scale 3 and rank 4. The mean of this feature was positively correlated with VAT over the entire data set (r=0.40). 
The mean and std of accel MSD for the 12-hour period were 0.22 and 0.14 for subject 1 and 0.15 and 0.09 for subject 2. The mean TDE eigenvalue was 1.93 for subject 1 and 2.22 for subject 2. So, the sign of the subject differences of the 12-hour summary statistics of these features is consistent with the global VAT correlations listed in Table~\ref{table:gait_features}. 

\begin{figure}[!t]
\centerline{
\includegraphics[width=1.75in]{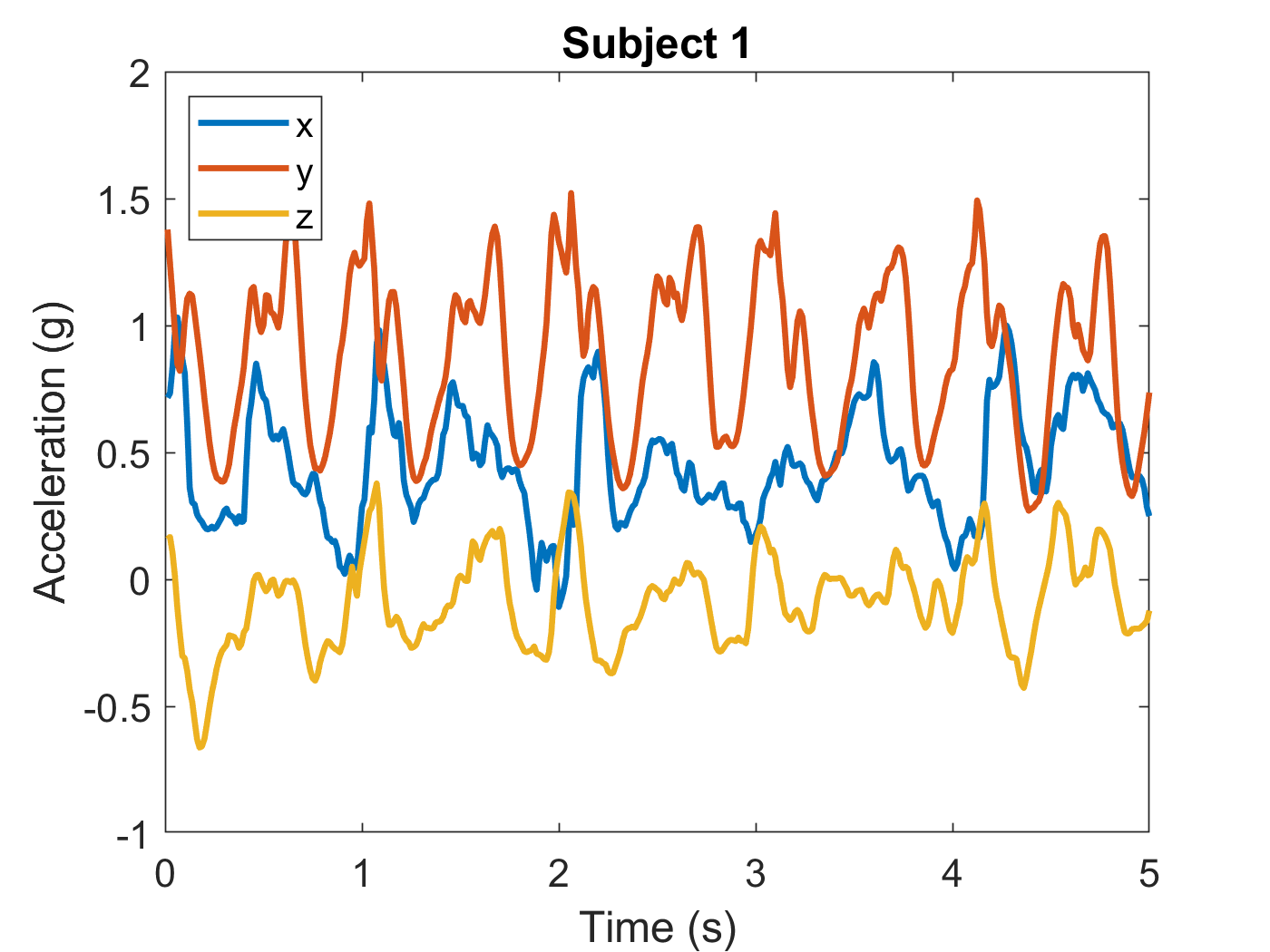}
\includegraphics[width=1.75in]{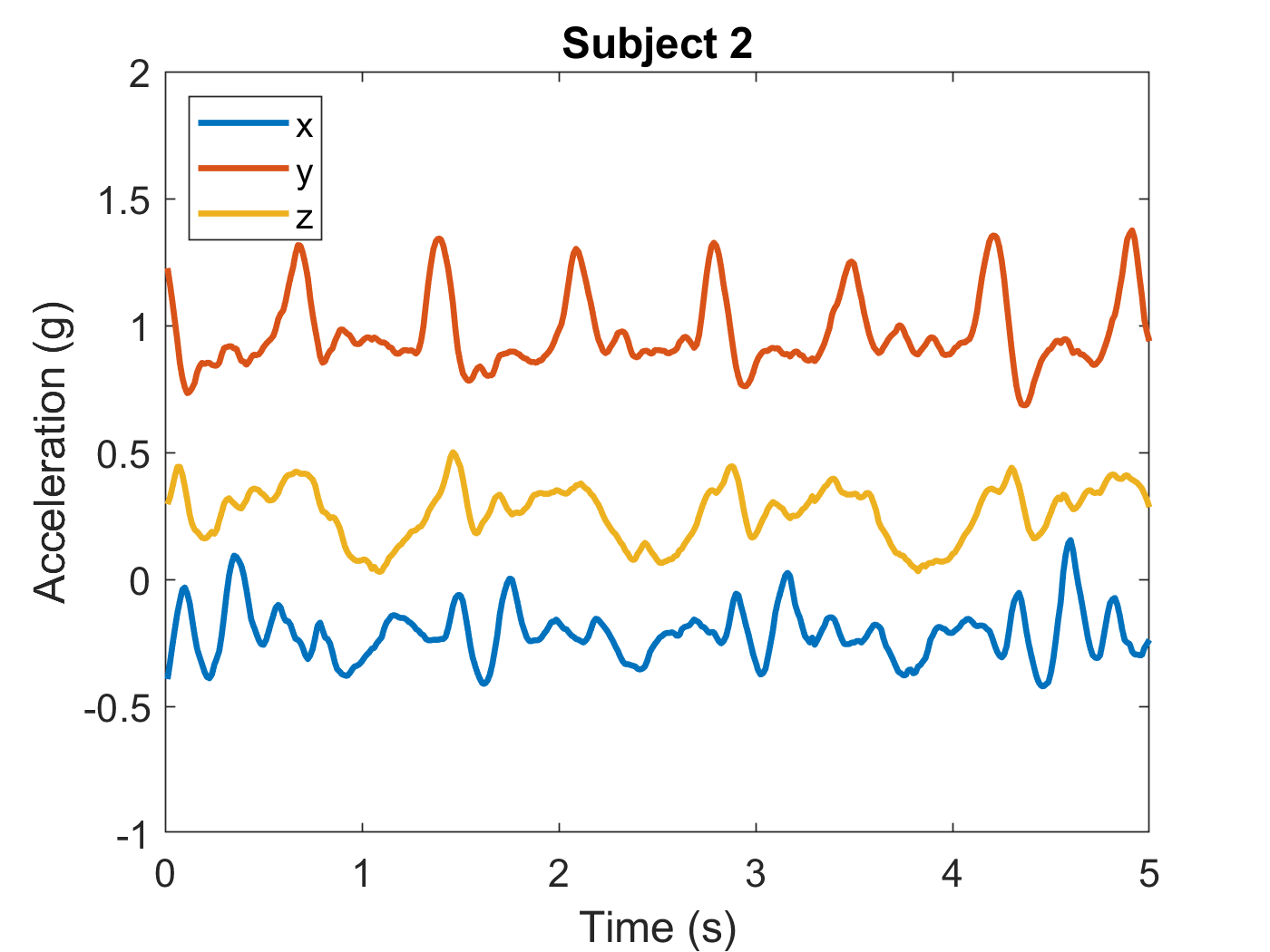}
}
\caption{Acceleration time series from a single 5 s gait frame for two example subjects. The selected gait frame is circled in red in Figure~\ref{fig:example_plot1}.
}
\label{fig:example_plot2}
\end{figure}

In order to provide an illustration of why the raw acceleration data might produce these feature differences, Figure~\ref{fig:example_plot2} plots the acceleration data from a single gait frame from each subject, highlighted by the red circles in Figure~\ref{fig:example_plot1}.
Subject~1 shows much larger magnitudes of acceleration than subject~2, presumably due to more vigorous arm swinging during gait. This explains why the accel MSD feature, which represented total acceleration magnitude variability during a frame, is an effective discriminant related to walking or running intensity.
The TDE eigenvalue features, however, did not contain any magnitude information, due to the fact that magnitude is factored out when computing the TDE correlation matrices. Why does subject~1 have a smaller eigenvalue at scale 3 and rank 4 than subject~2? This is due to the fact that vigorous arm swinging not only caused greater acceleration variability, it also caused a greater fraction of that total variability to be "explained" by dominant frequencies and phases associated with the arm swing. Therefore, a subject whose gait movements were characterized by less vigorous arm swings tended to have larger values in eigenvalues at slightly lower ranks (such as rank 4) because a larger fraction of the normalized time series variability is due to movements other the dominant arm swing movement. This effect was strongest at relatively large time delay scales that correspond to a step duration (scale 3) or stride duration (scale 4). It is notable that the average time delay for the TDE scale 3 time delays is 0.56 s, which is similar to the average step duration in the data set (0.59 s).
%

\subsection{VAT estimates from gait and sleep features}

Table~\ref{table:ridge_gait_sleep} shows the ability to estimate VAT with different feature configurations, in terms of the Spearman correlation of estimated VAT with measured VAT, as well as the mean absolute error. The mean and standard deviations of the results are shown across 30 test folds (based on 30 random train/test partitions with 0.8/0.2 splits).  For the gait features, two findings stand out. First, the features encoding gait patterns over time performed surprisingly poorly (r=0.283) relative to their correlations listed in Table~\ref{table:gait_features}. This was due to the fact that these features exhibit a surprising discrepancy in predictions by gender. In fact, within each gender, this feature class produced moderately strong correlations of r=0.358 (male) and r=0.296 (female). This level of gender-based discrepancy was not found for the other feature classes.

The second and more significant finding was that the gait dynamics features dramatically outperformed the other feature classes in ability to estimate VAT. This was almost entirely due to the TDE eigenvalue features, which by themselves produced VAT estimates with a correlation of r=0.631.

For the sleep features, two findings stand out. First, it was surprising that sleep movement intensity features actually achieve better VAT estimation accuracy than gait movement intensity features. Second, we see that the sleep movement dynamics features provided the strongest contribution (r=0.498). Again, the TDE eigenvalue features were mostly responsible, producing by themselves VAT estimates with r=0.490.
The best results were obtained by concatenating the gait and sleep feature vectors, resulting in a correlation with VAT of 0.702 and mean absolute error (MAE) of 151.9.

\begin{table}
\caption{VAT estimation based on gait and sleep features using ridge regression}
\begin{tabular}{|l|l|l|}
\hline
Features & VAT r & VAT MAE \\
\hline
1. Gait cadence \& regularity  & 0.300 $\pm$ 0.029& 204.9 $\pm$ 5.7\\
2. Gait intensity           & 0.346 $\pm$ 0.023& 204.8 $\pm$ 5.9\\
3. Gait patterns over time  & 0.283 $\pm$ 0.027& 208.0 $\pm$ 5.7\\
4. Gait dynamics            & 0.654 $\pm$ 0.020& 162.5 $\pm$ 3.9\\
\hline
5. Sleep \& movement durations  & 0.200 $\pm$ 0.034& 211.8 $\pm$ 5.4\\
6. Sleep movement intensity   & 0.401 $\pm$ 0.023& 196.1 $\pm$ 5.4\\
7. Sleep movement dynamics      & 0.498 $\pm$ 0.023& 185.2 $\pm$ 4.9\\
\hline
1-4 Gait all                     & 0.663 $\pm$ 0.022& 160.5 $\pm$ 4.1\\
5-7 Sleep all                    & 0.546 $\pm$ 0.022& 178.5 $\pm$ 5.0\\
1-7 Gait and sleep               & 0.702 $\pm$ 0.020& 151.9 $\pm$ 4.1\\
\hline
\end{tabular}
\label{table:ridge_gait_sleep}
\end{table}

\subsection{Comparison with MIMS features}

Table~\ref{table:mims_comparison} compares the ability to estimate VAT using the MIMS-derived features from \cite{williamson2023daily} with the gait and sleep features. 
Due to more stringent requirements on the accelerometry data, the approach in \cite{williamson2023daily} resulted in a smaller set of eligible subjects, 3,042 rather than 4,883. 
The MIMS-derived features consisted of two types: 1) features characterizing activity profiles over a day, and 2) features characterizing local-in-time activity fluctuations. The daily activity profile features were more effective, and the best MIMS-based result was obtained by fusing the two feature types. 
To make a fair comparison, the gait and sleep features were trained and tested on the smaller set of subjects, resulting in slightly lower accuracy than shown in Table~\ref{table:ridge_gait_sleep} (r=0.693 versus r=0.702). However, the gait and sleep features still produced VAT estimates that are far more accurate than the MIMS-based features.
Moreover, concatenating the gait and sleep feature vectors with the MIMS-based feature vectors reduced estimation accuracy compared to using gait and sleep features alone. 
These results provide strong evidence for the importance of deriving activity features from accelerometry at high time resolution as opposed to relying on summary activity statistics at 1-minute resolution, as provided by MIMS.

\begin{table}
\caption{Comparing MIMS features with Gait and Sleep features using ridge regression}
\begin{tabular}{|l|l|l|}
\hline
Gait Features & VAT r & VAT MAE (g)\\
\hline
1. MIMS daily activity profiles     & 0.367 $\pm$ 0.033& 204.4 $\pm$ 5.2\\
2. MIMS temporal activity fluct.    & 0.295 $\pm$ 0.030& 208.6 $\pm$ 4.6\\
1 \& 2.                             & 0.394 $\pm$ 0.030& 202.7 $\pm$ 4.8\\
3. Gait and sleep                   & 0.693 $\pm$ 0.015& 155.0 $\pm$ 3.8\\
1-3                                 & 0.679 $\pm$ 0.014& 159.8 $\pm$ 4.7\\
\hline
\end{tabular}
\label{table:mims_comparison}
\end{table}

\subsection{Deep Learning Transformers}

Due to computational constraints, the deep learning models were trained and tested on a single 70/10/20 train/validation/test split, in which the test data consisted of 969 subjects (489 male, 480 female).

As shown in Table \ref{table:transformer_results}, the performer network outperformed the basic network even though the performer network is not as deep. This is likely because the basic network's embedder acted as a bottleneck while the performer network's advanced attention mechanism is specifically designed to handle longer sequences. 
%
Using the MLP alone, trained with the 6-dimensional covariate variables, achieved accurate VAT estimates, with r=0.825. 
By combining acceleration features with covariates, the basic transformer achieved only slightly better estimates than using the covariates alone. However, the performer transformer achieved substantial improvement over the covariates alone, with r=0.856. Therefore, 
the activity-based time series information used by the transformer network provides complementary information to the static demographic and measurement information.

\begin{table}
\caption{Basic and Performer transformers}
\begin{tabular}{|l|l|l|}
\hline
Transformer Model & VAT r & VAT MAE (g) \\
\hline
Basic                   & 0.120 & 214.4 \\
Performer               & 0.668 & 159.0 \\
Covariates only         & 0.825 & 118.3 \\
Basic with covariates      & 0.835 & 123.9 \\
Performer with covariates  & 0.856 & 114.0 \\
\hline
\end{tabular}
\label{table:transformer_results}
\end{table}

%

\subsection{Fusing ridge regression and transformer VAT estimates}

Table~\ref{table:ridge_dnn_fusion} compares the accuracy of estimates using ridge regression and the performer transformer with different feature configurations. RMSE values are included to provide a comparison with the VAT standard deviations listed in Table~\ref{table:subject_measurements}. The first section of the table (1 and 2) shows results using 
accelerometry inputs alone. Here, ridge regression achieved moderately better results, and the best results were obtained by averaging the two approaches. 
The second section of the table (3 \& 4) shows results using covariate inputs alone. Here, the deep learning approach obtained moderately better results. A weighted average (1/3 ridge regression and 2/3 deep learning) produced a slight improvement. 
The third section (5 \& 6) shows results using accelerometry combined with covariate inputs.  Here, the deep learning approach achieved moderately better results in terms of correlation and MAE but not RMSE. A weighted average (1/3 ridge regression and 2/3 deep learning) produced an improvement in all measures. 

\begin{figure}[!t]
\centerline{
\includegraphics[width=.8\columnwidth]{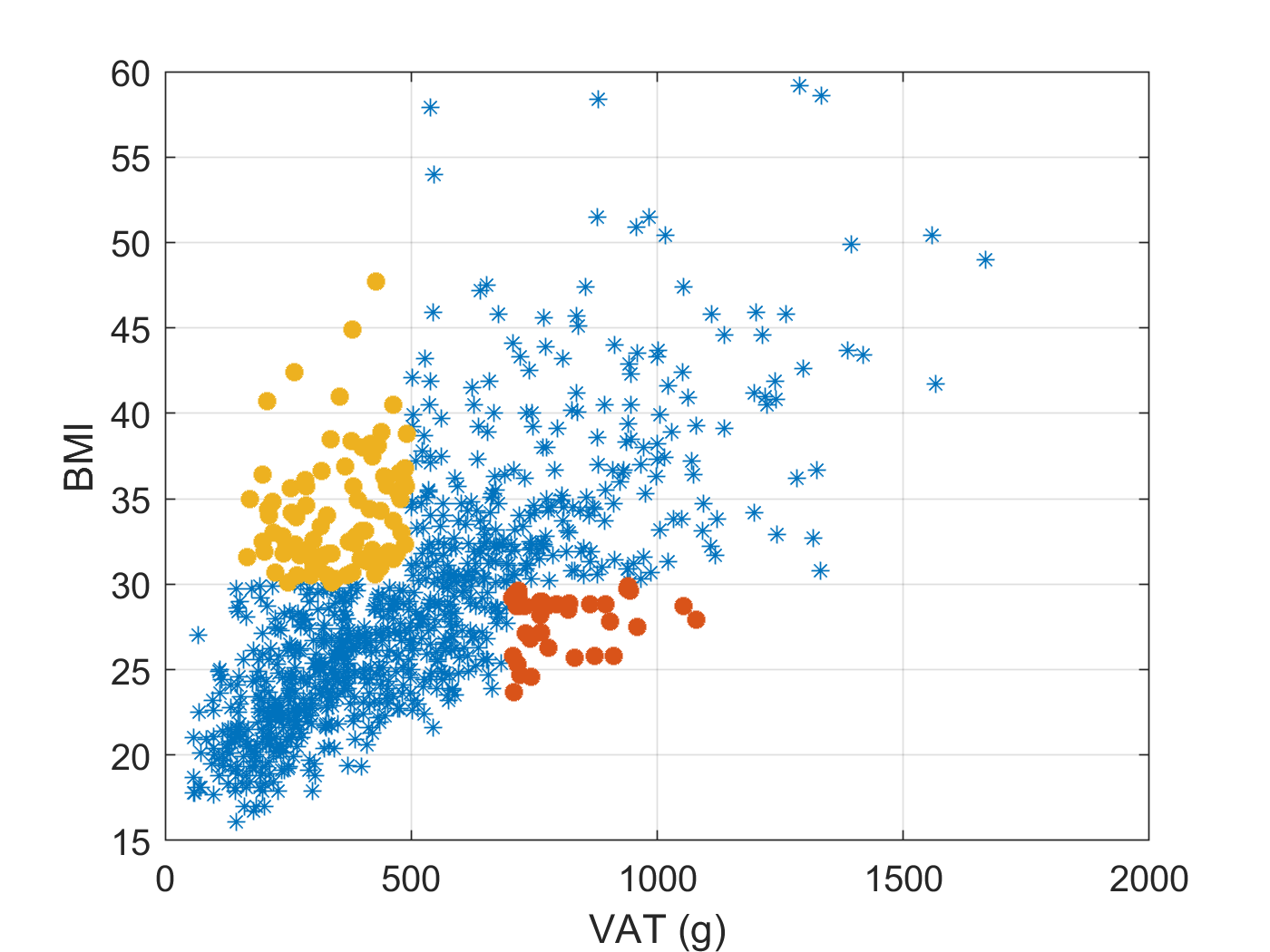}
}
\centerline{
\includegraphics[width=.8\columnwidth]{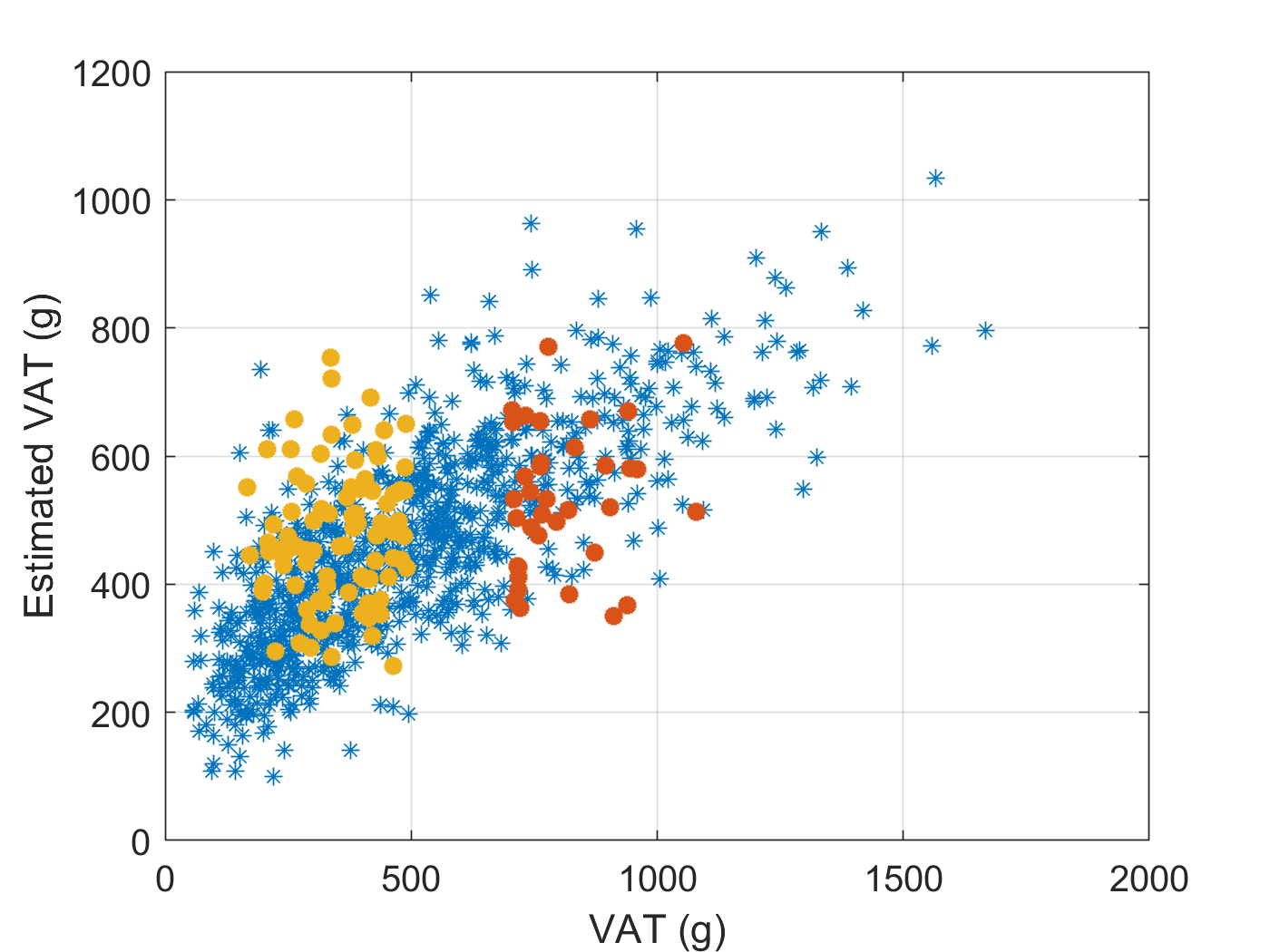}
}
\centerline{
\includegraphics[width=.8\columnwidth]{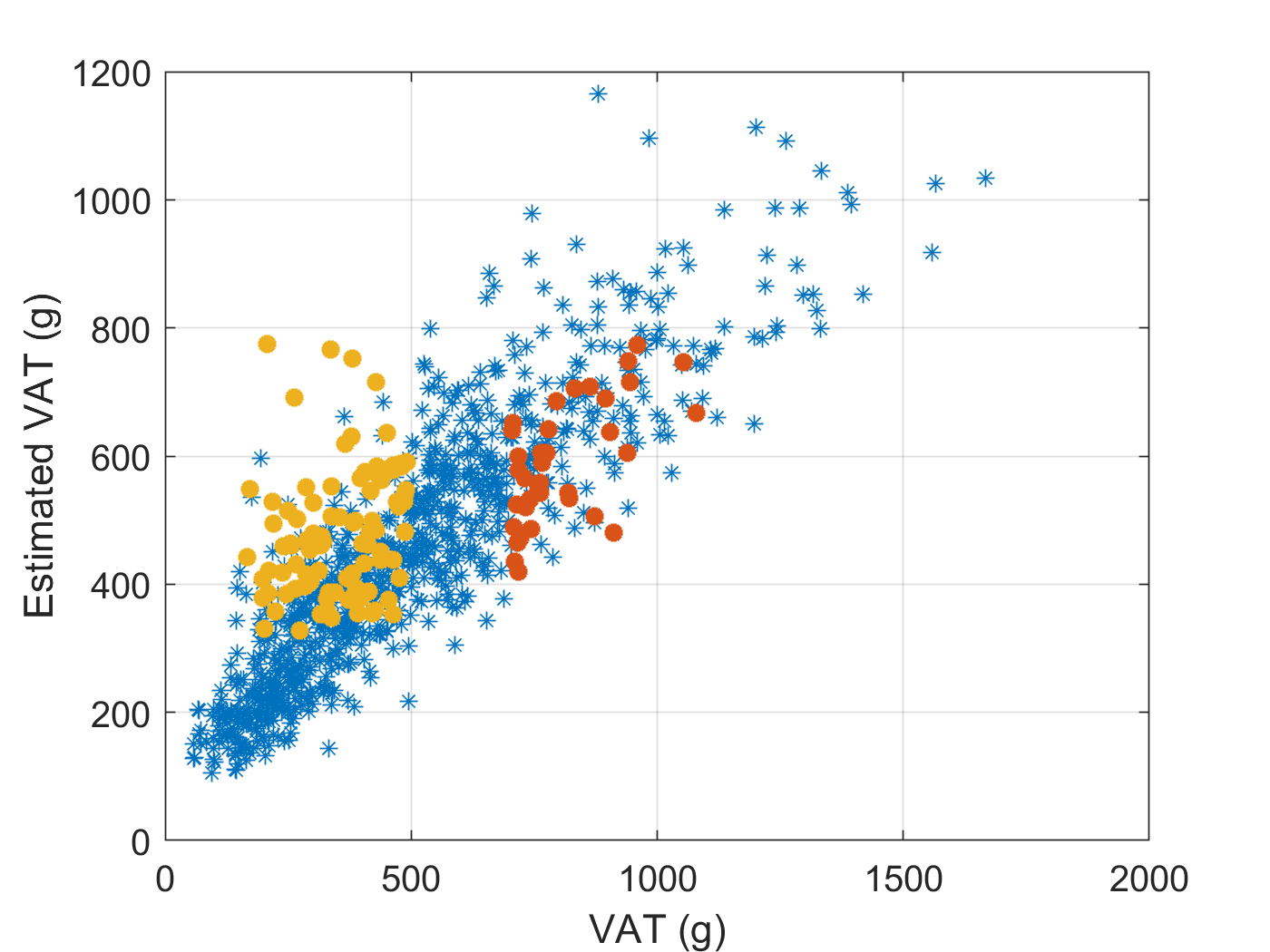}
}
\caption{Top: BMI is plotted as a function of measured VAT (r=0.69), with HB-LV subjects plotted in gold and LB-HV subjects in red (see text for details).
Middle: Estimated VAT based on activity alone as a function of measured VAT (r=0.73). Bottom: Estimated VAT based on activity and covariates as a function of measured VAT (r=0.86).}
\label{fig:vat_scatter}
\end{figure}

Figure~\ref{fig:vat_scatter} illustrates some advantages of using activity-based VAT estimates. Figure~\ref{fig:vat_scatter} (top) plots BMI as a function of VAT (r=0.69). It also plots subjects with high BMI but low VAT (HB-LV) values in gold and subjects with moderately low BMI but high VAT values (LB-HV) in red. Using BMI alone would classify the HB-LV subjects as being less healthy than the LB-HV subjects. 
Figure~\ref{fig:vat_scatter} (middle) plots VAT estimates based on activity as a function of VAT. Here, we see that activity-based estimates of VAT provide a better indications of metabolic health status for the HB-LV and LB-HV subjects.
Figure~\ref{fig:vat_scatter} (bottom) plots VAT estimates based on activity combined with covariates as a function of VAT (r=0.86). Here, we find much tighter VAT estimates overall, while also improving the  metabolic health indications for the HB-LV and LB-HV subjects.

%
%

%
%
%

Finally, Table~\ref{table:accuracy_by_bmi_categories} illustrates the value added by activity analysis over covariate information alone, using the best fused estimators. 
When testing on all subjects, the advantage of the fused estimator is moderate (r=0.829 for covariates alone, r=0.730 for acceleration alone, and r=0.858 for covariates and acceleration). We also divided the test population into the standard BMI-based classes (normal, overweight and obese). We found that when testing is restricted to overweight and/or obese subpopulations, the accuracy increment from adding accelerometry to the estimation models becomes larger. 

\begin{table}
\caption{VAT estimation using ridge regression and DNN models}
\begin{tabular}{|l|l|l|l|}
\hline
Features & r &MAE (g) & RMSE (g) \\
\hline
1. Gait \& sleep (ridge) & 0.696 & 150.9 & 197.6 \\
2. Performer Transformer    & 0.668 & 159.0 & 213.7 \\
1. \& 2.                    & 0.730 & 145.3 & 193.2 \\
\hline
3. Covariates only (ridge)  & 0.808 & 124.9 & 165.1 \\
4. Covariates only (MLP)  & 0.825 & 118.3 & 164.9 \\
3. \& 4.                    & 0.829 & 117.0 & 159.8 \\
\hline
5. Gait \& sleep with cov. (ridge)    & 0.825 & 118.8 & 154.8 \\
6. Performer with covariates    & 0.856 & 114.0 & 159.5 \\
5. \& 6.                    & 0.858 & 109.5 & 150.6 \\
\hline
\end{tabular}
\label{table:ridge_dnn_fusion}
\end{table}

\begin{table}
\caption{VAT correlations as function of BMI category}
\begin{tabular}{|l|l|l|l|}
\hline
BMI category & Cov. & Accel. & Cov. \& Accel \\
\hline
All (n=969)                         & 0.829 & 0.730 & 0.858 \\
Normal (n=304)            & 0.808 & 0.599 & 0.822 \\
Overweight or Obese (n=665)& 0.722 & 0.617 & 0.775 \\
Overweight (n=298)& 0.636 & 0.465 & 0.712 \\
Obese (n=367)          & 0.677 & 0.600 & 0.734 \\
\hline
\end{tabular}
\label{table:accuracy_by_bmi_categories}
\end{table}


\section{Discussion}

This paper provides novel contributions to the study of how activity is associated with health outcomes by using models trained from wrist-worn accelerometry to estimate VAT, an important marker of human health.
While daily activity and VAT have been previously associated, this study provided a detailed exploration of aspects of daily human activity and demonstrated the value added from this deeper investigation of the higher frequency accelerometry compared to one minute summary data.  

There is a previous study with many comparable aims, in which activity measurements were combined with demographic and body measurements in a regression model to predict VAT and liver fat, which is another indicator of metabolic risk \cite{keating2016objectively}. The subject population in that study was restricted to overweight or obese people who reported exercising $\leq$ 3 days a week and did not meet current physical activity guidelines. The activity features used in \cite{keating2016objectively} were derived from an accelerometer worn on the upper arm, rather than wrist, and the accelerometry features were levels of total sedentary, light, moderate, and vigorous physical activity. These are features that can be derived from low time resolution summary statistics. Several additional covariates were also used that are not included in our study: systolic and diastolic blood pressure, triglycerides, alanine
aminotransferase, glucose, and insulin. 

The only activity measure in \cite{keating2016objectively} that had a sizeable correlation with VAT was moderate activity, which correlated with $r=0.294$. This correlation has a similar magnitude to the correlations of movement intensity and VAT shown in our paper in Table 2, but surprisingly has the opposite sign. The sign of movement intensity correlations in our study is more consistent with other findings in the literature. There is discussion in \cite{keating2016objectively} for why this correlation has the opposite sign than what is expected. One possible explanation is the subject selection criteria, which restricted the study to a physically inactive subpopulation.

In any case, as shown in Tables 5-7 in our paper, the strongest activity-based correlations with VAT did not involve movement intensity features at all, but rather movement dynamics features, which require high time resolution acceleration data. These included  gait dynamics features (r=0.654) and sleep movement dynamics features (r=0.498). In addition, the deep learning model (r=0.668) uses features from a foundation model that needs high time resolution (30 Hz) acceleration data to generate the features.

The closest comparison to the model of \cite{keating2016objectively} is to restrict our test results to an overweight and obese subpopulation and use acceleration data only. When we do this, our model achieved much higher correlations with VAT, as shown in Table~\ref{table:accuracy_by_bmi_categories} (r=0.617 versus r=0.286).

Another interesting comparison with \cite{keating2016objectively} is to assess the amount that the activity features contribute to the prediction accuracy above what is obtained using covariates alone. 
In \cite{keating2016objectively}, with the covariates alone the regression model produced a global correlation with VAT of r=0.790. Adding the activity features boosted this only slightly to r=0.808. In our study, using the subpopulation of overweight and obese subjects, with covariates alone we obtained a VAT correlation of r=0.722. Our lower correlation is presumably because we did not include the multiple blood pressure and biochemical measurements used in \cite{keating2016objectively}.  When we added the activity features to the regression model, we obtained a much larger boost in accuracy, with an increase from r=0.722 to r=0.775.

The analysis of a wide range of activity features in our paper opens the door to further investigations of the key aspects of human activity that have the greatest influence on VAT.  This could range from the distribution of activity patterns throughout the day, the intensity of activities, to the types of activity (e.g., non-exercise activity thermogenesis, purposeful exercise, and patterns of physical labor).  This study also revealed that easily measured personal characteristics such as height, weight and waist circumference can further improve the activity factor prediction of VAT.  Other ongoing studies are attempting to define clinically meaningful thresholds of VAT which will provide specific targets for activity-based medical management, including recent work currently under review \cite{potter2025visceral}.

More specifically, the first modeling approach approach extracted movement features during gait and sleep, demonstrating most prominently the effectiveness of time-delay embedding for analysis of movement dynamics. The second modeling approach combined a DNN foundation model that characterized local acceleration patterns, with a DNN performer transformer model that derived VAT estimates from the temporal patterns throughout a single day. With these two approaches, the paper demonstrated the usefulness of using high time resolution acceleration data as the basis for health monitoring, and the usefulness of combining activity-based estimation with subject-level covariates such as gender, age, height, weight, and waist circumference. By combining both activity modeling approaches with subject covariates, the model produced VAT estimates with high correlation (r=0.858) and low mean absolute error (109 g) and root mean squared error (151 g). Finally, the contribution from activity analysis over the use of covariates alone was quantified, with the greatest added value provided for overweight and obese subjects. 




One interesting finding is that the DNN performance improved more when  covariates were combined with activity features. This may be due to the more powerful learning capabilities of the MLP component of the DNN pipeline, compared to ridge regression.
However, it remains a promising finding that the engineered activity features, conditioned on the specific behaviors of gait and sleep, performed competitively with the DNNs. 
The engineered feature approach offers the potential of providing greater insight and transparency. The approach could also be extended by applying it to other specific behavior classes, and by finding better ways to leverage the covariate information during regression modeling.

The engineered feature approach also has the advantage of connecting to extensive previous research into how aspects of gait and sleep, measured from accelerometry, relate to obesity in general and VAT in particular. The feature-VAT correlations, summarized in tables~\ref{table:gait_features} and \ref{table:sleep_features}, are consistent with previous findings in terms of the directionality of correlations.

One surprising finding was that, for features extracted during gait especially but also during sleep, the TDE movement dynamics features were the most predictive of VAT. This suggests that research in activity-based estimation of health outcomes could be improved by taking advantage of the high time resolution and signal to noise ratio provided by accelerometry and by focusing on more information-rich features that quantify movement time-frequency and dimensionality properties. These features could provide an effective complement to more standard measurements such as step counts, cadences, activity counts, or average activity levels.

Limitations of the current study primarily concern the interpretability of results. 
Wrist-worn accelerometry-based activity analysis demonstrated aspects of movement that correlate with VAT. The direction of causation is indeterminate. Part of the correlation is undoubtedly due to the fact that being overweight or in poor health causes individuals to move differently. Therefore, the prescriptive utility of the findings is not clear, in terms of the degree to which recommendations for changing activity will improve health outcomes. 
Another limitation is due to the fact that the data was collected in free-living conditions, and so inferences about the meaning of correlations of gait and sleep features must be constrained by lack of certainty about the accuracy with which the underlying behaviors were classified from the wrist-worn accelerometry.

In future work we will explore the analysis of other behavior classes, in addition to gait and sleep, for focused exploration of activity features. Future work will also include the analysis of additional subject covariates and health outcomes to enable a richer analysis of how multiple health factors relate to each other and to physical activity. Finally, we plan to analyze the DNN model to gain greater insight into the meaning of the foundation model features, and how temporal patterns of these features correlate with VAT.





\section*{Disclaimers}
The opinions and assertions in this manuscript are solely those of the authors and do not necessarily represent the official views or policies of the US Department of Defense, US Army, or US Air Force…

\bibliographystyle{IEEEtran}

\bibliography{refs}

\end{document}